\documentclass[twoside,a5paper]{article}
\usepackage[utf8]{inputenc}
\usepackage[voffset=1.5mm,top=15mm,height=17.9cm,left=2cm,width=10.8cm]{geometry}
\usepackage{fancyhdr}
\usepackage{ngerman}
\usepackage{sectsty}
\usepackage[T1]{fontenc}
\usepackage{tgtermes}
\usepackage{microtype}
\usepackage{semtrans}
\usepackage{graphicx}
\usepackage{footmisc}

\fancypagestyle{firstpage}
{
	\fancyhf{}
	\fancyhead[L]{Mitteilungen der Deutschen Orient-Gesellschaft zu Berlin}
	\fancyhead[R]{141 $\cdot$ 2009}
	\fancyfoot[LE,RO]{\thepage}
}
\allsectionsfont{\normalsize\centering\mdseries\itshape}

\newenvironment{bib}{\fontfamily{ptm}\fontsize{9}{12}\selectfont}{}
\newenvironment{figs}{\footnotesize}{}

\begin{document}
\setcounter{page}{45}
\vspace*{2.5cm}
\begin{center}
\large{\textbf{Geometrie und Astronomie im Stadtplan\\ des hethitischen
Sarissa}} \vspace*{5mm}

\normalsize{\textsc{Andreas Müller-Karpe/Vuslat Müller-Karpe/Andreas 
Schrimpf\footnote{Vorgeschichtliches Seminar der Philipps-Universität Marburg, 
e-mail:  andreas.muellerkarpe@staff.uni-marburg.de, 
vuslat.muellerkarpe@staff.uni-marburg.de, bzw.\ 
Fachbereich Physik der Philipps-Universität Marburg, 
e-Mail: andreas.schrimpf@physik.uni-marburg.de.}}}
\end{center}

\thispagestyle{firstpage}

\section*{Zusammenfassung}
Die hethitische Stadt Sarissa wurde im 16.\ Jh.\ v.\ Chr.\ gegründet und
planmäßig ausgebaut.
Die Stadttore und die größeren öffentlichen Gebäude sind um etwa 45$^\circ$ gegen die 
Himmelsrichtungen gedreht angeordnet. Ein Tempel jedoch weist eine deutliche Abweichung 
davon auf. Die folgende Analyse zeigt, dass bei beiden Orientierungen astronomische Bezüge, 
insbesondere der Sonnenlauf, von Bedeutung waren, ja dass mit gewisser Wahrscheinlichkeit 
auch ein Bezug zu markanten Punkten der Venus eine Rolle gespielt haben mag.

\section*{Die Ausrichtung der Stadt und des Tempels 1 auf der Nordterrasse}
Als planmäßig angelegte Neugründung unterscheidet sich die hethitische Stadt Sarissa von der 
Mehrzahl der anderen bislang bekannten Orte Altanatoliens. Meistens sind Städte aus alten 
Siedlungskernen heraus gewachsen und entsprechend unregelmäßig in ihrem Aufbau. 
Anders Sarissa. Hier konnten während 12 Grabungskampagnen (von 1993-2004) keine 
Hinweise auf eine vor- oder frühhethitische Besiedlung nachgewiesen 
werden\footnote{Müller-Karpe 2009a mit Hinweisen zu älterer Literatur; 
zu den Rahmenbedingungen der Gründung Müller-Karpe 2009b.}. 
Alles spricht dafür, dass die Stadt ohne älteren Vorgänger gewissermaßen ,,am Reißbrett'' 
entworfen wurde. Der Grundplan nimmt dabei deutlich Bezug auf die Himmelsrichtungen, 
wobei allerdings auch lokale topographische Gegebenheiten geschickt in diesen Plan 
integriert wurden (Abb.~1).

\begin{figs}
\begin{center}
\includegraphics[angle=0,width=0.99\textwidth]{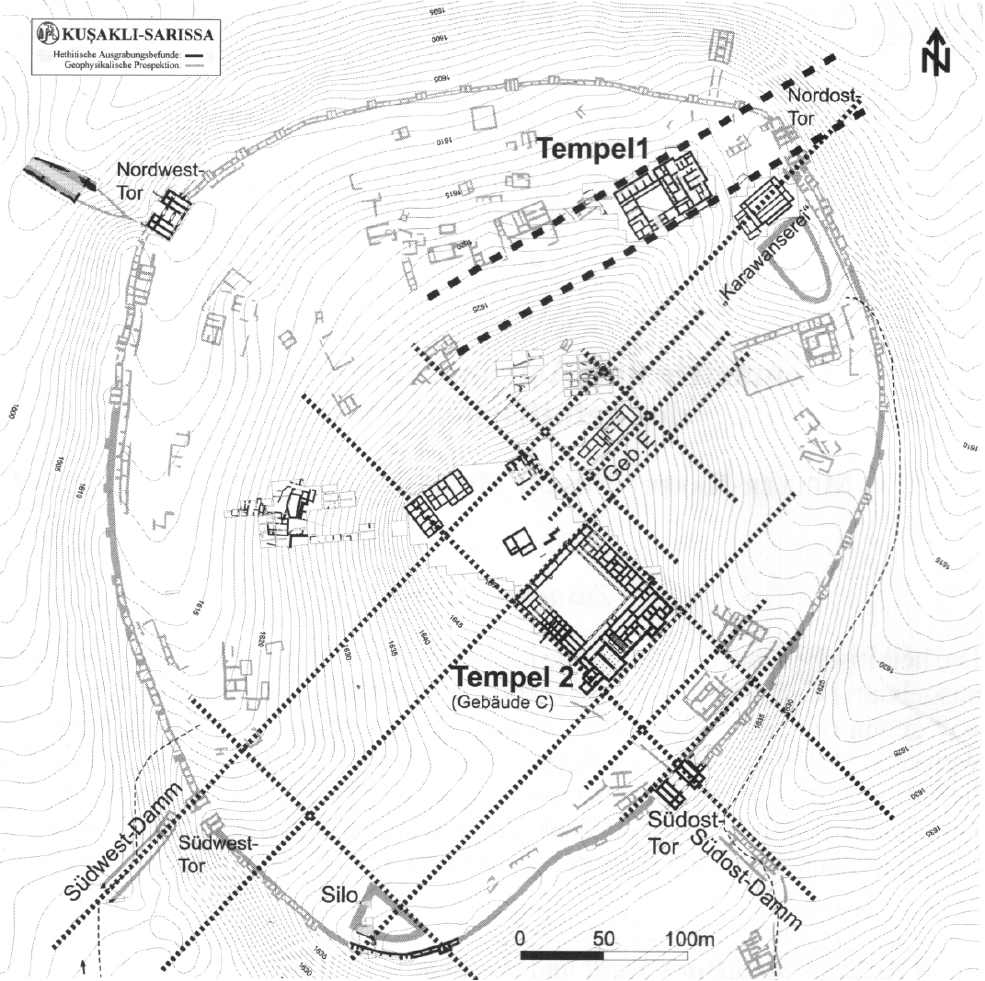}

\vspace*{1em}
Abb.\ 1: Plan der hethitischen Stadt Kuşaklı-Sarissa in Ost-Kappadokien mit Eintragung der 
Planungsachsen aus der Gründungsphase (letztes Drittel des 16.\ Jh.\ v.\ Chr.). Die 
Ausrichtung des Tempels 1 differiert signifikant gegenüber den übrigen öffentlichen Bauten.
\end{center}
\end{figs}

\vspace*{1em}

Die vier Stadttore sind jeweils nach Nordwest, Nordost, Südwest und Südost ausgerichtet. 
Hierdurch kommt bereits zum Ausdruck, dass nicht die einfachen Himmelsrichtungen 
(Nord, Süd, West, Ost) primäres Bezugssystem für die Planungen war, sondern die Diagonalen. 

Zieht man eine Linie von dem Nordosttor zu dem Südwesttor, so trifft diese genau im 
rechten Winkel auf die Verlängerung der Torgasse durch das Südosttor. An diesem 
Achsenkreuz ist der Große Tempel am Südosthang der Akropolis (Tempel 2 bzw.\ Gebäude C) 
orientiert\footnote{Müller-Karpe 2000.}. Er ist der größte und ehemals zweifellos auch 
der bedeutendste Bau der 
Stadt gewesen. Vermutlich wurde der ,,Wettergott von Sarissa'' hier verehrt\footnote{Zur 
Bedeutung dieses Wettergottes siehe Wilhelm 1997,14 f.}. Aufgrund 
archäologischer Belege ist die Errichtung dieses zentralen Sakralbaus wie auch der 
Stadttore (damit
\newpage

\begin{figs}
\begin{center}
\includegraphics[angle=0,width=0.99\textwidth]{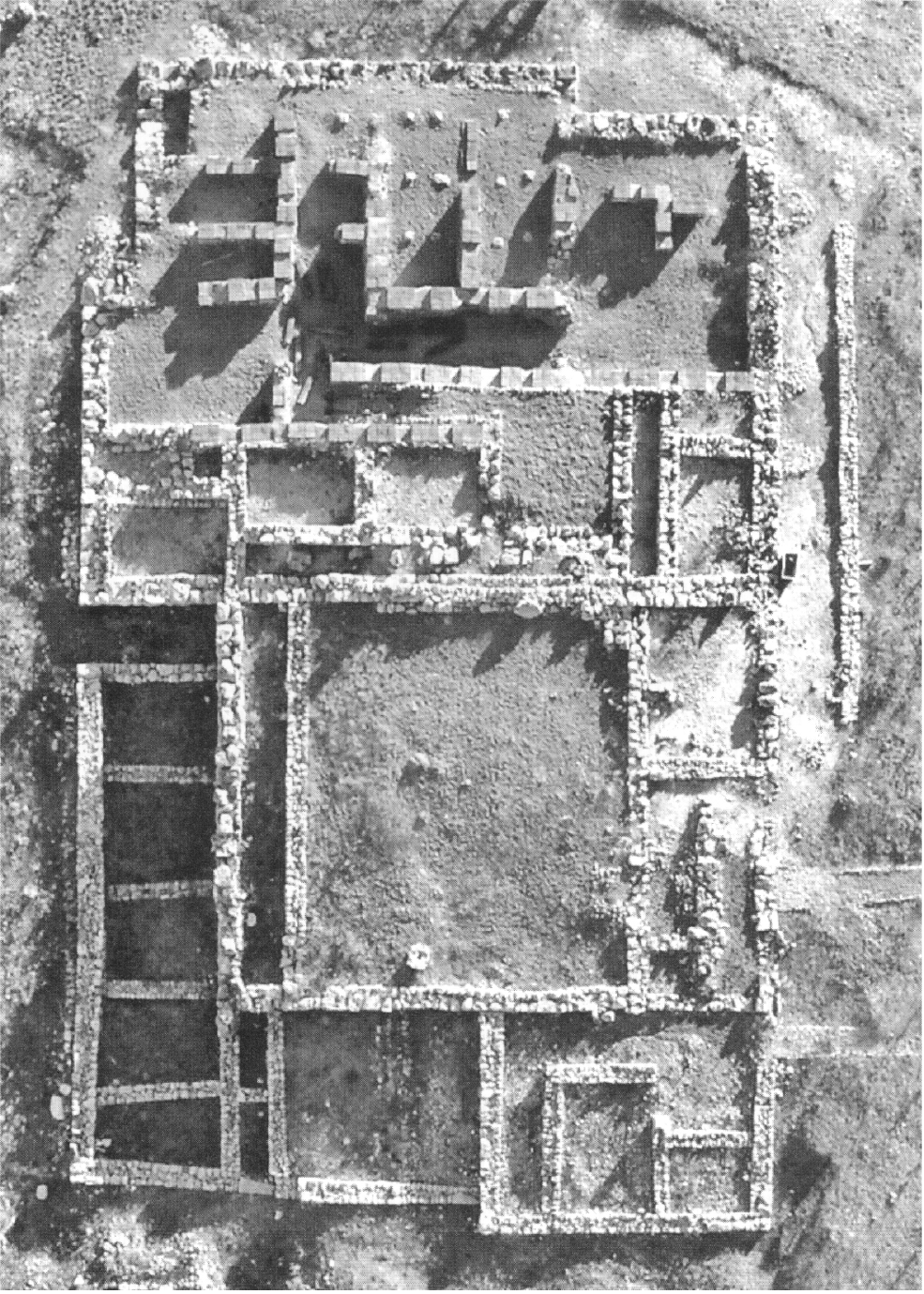}

\vspace*{3em}
Abb.\ 2: Luftaufnahme des Tempels 1 auf der Nordterrasse von Kuşaklı-Sarissa,\\ 
des potentiellen Ištar bzw.\ Anzili-Tempels.
\end{center}
\end{figs}

\newpage
\noindent
die gesamte Stadtbefestigung) in althethitische Zeit zu datieren. 
Ergebnisse dendrochronologischer Untersuchungen an verkohlten Bauhölzern sowohl des 
großen Tempels, wie auch der Tore, vermögen den Errichtungszeitraum auf die 30er/20er 
Jahre des 16.\ Jh.\ v.\ Chr.\ einzuengen\footnote{Kuniholm/Newton 1996.}.

\vspace{1em}
\noindent
Doch nicht nur die genannten Anlagen gehören zu dieser Gründungsplanung der Stadt. Die 
selbe Orientierung wie der große Tempel weist das nur wenig nördlich gelegene Gebäude E auf, 
sowie die ca.\ 30~m lange nordöstliche Front des D-förmigen Getreidespeichers auf der 
Südspitze. Auch die sogenannte Karawanserei auf der Innenseite des Nordosttores zeigt diese 
Ausrichtung. Gleiches gilt für die vier außerhalb der Stadt vor den Toren errichteten 
Staudämme.

Hingegen lässt sich bei der einfachen Wohnbebauung diese Orientierung nicht nachweisen. 
Es deutet sich somit an, dass primär die großen öffentlichen Bauten dem ,,Masterplan'' 
gemäß errichtet wurden. Dieser Plan ist gegenüber der Nordlinie um ca. 45$^\circ$ nach Osten 
hin gedreht\footnote{Der genaue Wert ist 44,3$^\circ$. 
Die Bestimmung der geographischen Ausrichtung erfolgte am 15.8.2009 nach Sonnenuntergang 
mit Hilfe eines Theodolithen durch das Kayalıpınar Grabungsteam (K.\ Bieber, T.\ Brestel, 
A.\ u.\ V.\ Müller-Karpe, Chr.\ Salzmann, K.\ Sauer, J.\ Wangen). Es wurde hierbei der 
Winkel zwischen der südöstlichen Außenmauer des Tempels 1 und dem Nordstern (Polaris) 
sowie weiteren Gestirnen gemessen. Diese Werte wurden dann von A.\ Schrimpf mit Hilfe 
des ,,Horizons''-Programms der NASA umgerechnet. Zu bedenken ist allerdings, dass 
die als Basislinie der Messung verwendeten Fundamentsteine der Tempelmauer keine exakte 
Kante bilden, da es sich hier nicht um Quadermauerwerk handelt. Die Werte mehrerer, 
jeweils geringfügig divergierender Messungen wurden gemittelt.}.
 Man wählte somit die Mitte zwischen der Nord-Süd- und der West-Ost-Richtung. 

Auffällig ist aber, dass ein bedeutender Bau der Stadt eine deutlich ab-weichende 
Orientierung aufweist, obwohl er bereits aus althethitischer Zeit stammt: Der Tempel 1 
auf der Nordterrasse (Abb.~2). Auch nach den dendrochronologischen Untersuchungen ist 
dieser Bau nicht jünger als der große Tempel. Seine Längsachse liegt im Winkel von 59,1$^\circ$ 
zur Nordlinie (Abb.~3). Im Gelände fällt zwar die Abweichung von knapp 15$^\circ$ gegenüber 
den ,,Stadtplanungsachsen'' nicht auf, angesichts der erstaunlichen Präzision, mit 
der die übrigen öffentlichen Bauten der Gründungsphase errichtet wurden, ist die 
Abweichung im Stadtplan bezüglich der Positionierung des Tempels auf der Nordterrasse 
doch bemerkenswert.

In der hethitischen Hauptstadt gefundenen Bauritualen ist zu entnehmen, dass bei der 
Gründung eines Tempels wohl nichts dem Zufall überlassen wird. Akribisch genau werden 
die Details der Zeremonien anlässlich der Grundsteinlegung der Bauten 
geschildert\footnote{Götze 1950; Haas 1994: 252-256.}. 
Auch wenn dies in den überlieferten Texten nicht explizit dargelegt wird, so ist doch 
davon auszugehen, dass gleichwohl die Orientierung eines neu zu errichtenden Tempels 
gut überlegt gewesen sein dürfte. Dass hierbei nicht allein praktische Gesichtspunkte 
(Geländeform, Hauptwindrichtung etc.) sondern Aspekte des in dem Gebäude auszuübenden 
Kultes, somit auch eine symbolische Bedeutung, eine Rolle gespielt haben dürfte, 
kann bei einem Sakralbau vorausgesetzt werden. 

\begin{figure*}[t]
\begin{figs}
\begin{center}
\includegraphics[angle=0,width=0.99\textwidth]{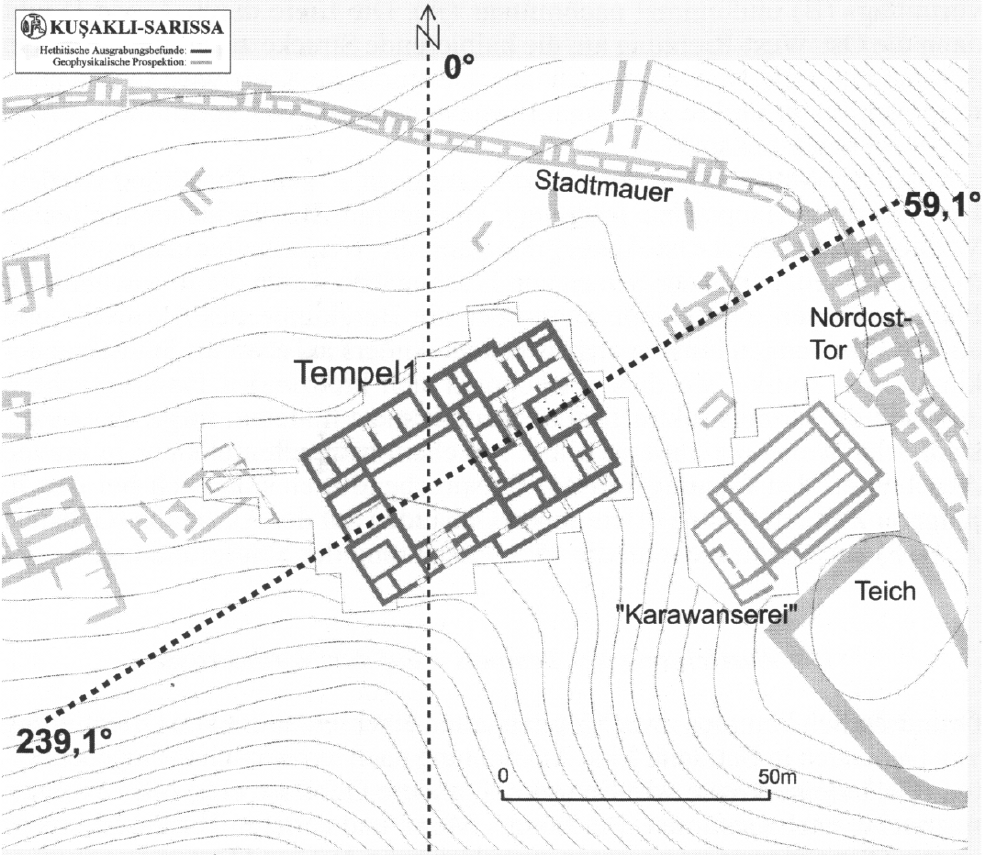}

\vspace*{0.5em}
Abb.\ 3: Plan der Nordterrasse von Sarissa mit Ausrichtung der Längsachse\\ des Tempels 1 
in Relation zur Nordlinie.
\end{center}
\end{figs}
\end{figure*}

\section*{Die grundsätzliche Ausrichtung der Tore und der Stadt}

Wie aber konnte man in hethitischer Zeit diesen 45$^\circ$-Winkel bestimmen und warum tat man 
dies? Voraussetzung war zunächst die Bestimmung der Haupthimmelsrichtungen. 
Hierzu gab es prinzipiell zwei Möglichkeiten. Entweder nachts anhand der Drehung 
der Sterne um die Nordrichtung oder tags mit Hilfe des Schattenwurfs der Sonne\footnote{Zu 
den Methoden der Bestimmung der exakten Nordrichtung bei den Ägyptern siehe 
Stadelmann 1992, 118 f.}. 
Eine zweifellos bereits den Hethitern bekannte Methode der Bestimmung der 
Haupthimmelsrichtungen ist der sogenannte Indische Kreis. Hierbei schlägt man um einen 
senkrecht im Boden stehenden Stab einen Kreis, dessen Radius etwas größer ist als 
die Länge des mittäglichen Schattens. Dann schneidet die Spitze des Stabschattens 
diesen Kreis einmal vormittags (B) und einmal nachmittags (A). Die Linie durch A und B 
gibt dann die Ost-West-Richtung an, die halbierende Strecke zusammen mit dem Stabfußpunkt 
die Nord-Süd-Richtung. Der Indische Kreis wird in manchen Regionen bis in 
unsere Zeit hinein angewandt und war vermutlich bereits in der Steinzeit 
bekannt\footnote{Schlosser/Mildenberger/Reinhardt/Cierny 1997:\ 57.}.

Mit dieser Methode kann die Nordrichtung auf 2$^\circ$ genau bestimmt werden. Hat man diese 
Hauptachse festgelegt, können mit Zirkelschlägen der rechte Winkel hierzu (West-Ost-Achse) 
und dann die Winkelhalbierende (45$^\circ$) ermittelt werden. Durch das Setzen von Fluchtstäben 
waren die so gewonnenen Planungsachsen im Gelände auch über die Bergkuppe der Akropolis 
und andere Erhebungen hinweg auszustecken. Anders als etwa im Städtebau der klassischen 
Antike, wo diese orthogonal sich kreuzenden Planungsachsen zugleich die Grundstruktur 
für das Verkehrsnetz bildeten, folgte das innerstädtische Wegenetz Sarissas allem 
Anschein nach allenfalls nur in kurzen Abschnitten den Planungsachsen. Entsprechend 
hielt man sich bei den in jüngerer Zeit errichteten öffentlichen Gebäuden auf der 
Akropolis (Gebäude A, B, F) auch nicht mehr an die 45$^\circ$-Ausrichtung der Stadtgründung.

\section*{Die Ausrichtung des Tempels 1 auf der Nordterrasse}

Grundsätzlich kommen zweierlei Möglichkeiten einer intentionellen Orientierung von 
Gebäuden in Betracht: Eine Ausrichtung nach terrestrischen Gegebenheiten 
(Geländemerkmale wie Berggipfel, andere Gebäude etc.) oder extraterrestrische, 
nach Gestirnen. Bei Sonne, Mond und anderen Himmelskörpern unseres Sonnensystems 
kommen jeweils nur die Auf- und Untergangspunkte am Horizont für eine 
Gebäudeorientierung in Frage. All diese Himmelskörper liegen mit nur geringen 
Abweichungen in einer Ebene, der Ekliptik, die zurzeit um etwa 23,5$^\circ$ gegenüber 
der Äquatorebene geneigt ist. Aufgrund des jährlichen Umlaufs der Erde um die 
Sonne tauchen diese Körper im Laufe eines Jahres zyklisch um bis zu etwa 23,5$^\circ$ 
über und unter dem Äquator auf. Wenn es keine ausgezeichneten Kalendertage gibt, 
die für die Erbauer der Stadt von besonderer Bedeutung gewesen sind, ist die 
Wahrscheinlichkeit groß, dass die geographischen Extreme, die nördlichen und 
südlichen Wenden der Auf- und Untergangspunkte eine Rolle gespielt haben könnten.

Für das hethitische Sarissa ist den bisher gefundenen Texten zufolge der Kult weder 
einer Sonnen- noch einer Mondgottheit belegt. Es wurde jedoch nachweislich eine 
Ištargestalt (hier wohl Anzili genannt\footnote{Primär aufgrund eines in Kuşaklı-Sarissa 
gefundenen Textes konnte G.\ Wilhelm überzeugend belegen, dass sich hinter der 
im hethitischen Schrifttum geläufigen Schreibung \textsl{IŠTAR-li} die Lesung ,,Anzili'' 
(bzw.\ ,,Enzili'') verbirgt: G.\ Wilhelm 2002: 345 und Wilhelm im Druck.}) verehrt, für die ein Tempel vorhanden 
gewesen sein muss. Ištar aber und der Planet Venus sind untrennbar miteinander 
verbunden\footnote{Haas 1994, 351.}. Neben dem Wettergott als dominierender Gottheit der Stadt und einer 
weniger wichtigen ,,Schutzgottheit'' ($^{\mbox{\tiny D}}$KAL) dürfte somit die genannte Ištargestalt 
wohl den zweiten Platz im Kult Sarissas eingenommen haben. Es mag daher legitim 
sein, die Arbeitshypothese zu formulieren, dass ihr der zweitgrößte Sakralbau 
der Stadt, der Tempel 1 auf der Nordterrasse, zuzuordnen ist (Abb.~2). Gestützt 
wird diese These auch durch den Befund, dass von allen 31 bisher in Boğazköy 
ausgegrabenen hethitischen Sakralbauten der Tempel 7 an der Sarıkale die größten 
Übereinstimmungen bezüglich der Grundrissgestaltung mit dem Tempel auf der 
Nordterrasse in Kuşaklı zeigt und Tempel 7 ist der einzige, in dem ein Ištarfigürchen 
gefunden wurde\footnote{Zum Baubefund des Tempels 7: Neve 1999,34-45, Taf. 25 c 
(Ištar-Figürchen). Allerdings nimmt Neve keine Zuweisung zu einer bestimmten Gottheit 
vor.}. Bereits bald nach der Ausgrabung kam daher der Gedanke auf, ob nicht zwischen der 
Venusbahn und der Ausrichtung des Tempels 1 auf der Nordterrasse von Sarissa eine 
Verbindung bestanden haben könnte. 

Die Venus ist der von der Erde aus nächst innenliegende Planet in unserem Sonnensystem. 
Sie ist daher nur in Blickrichtung zur Sonne zu beobachten, niemals nachts mit der 
Sonne im Rücken. Sie erscheint uns als Morgenstern, wenn sie vor der Sonne aufgeht, 
und als Abendstern, wenn sie nach der Sonne untergeht. Der Abstand zwischen Venus- und 
Sonnenaufgang oder Sonnen- und Venusuntergang kann jeweils mehr als vier Stunden betragen.

Für einen Umlauf um die Sonne benötigt die Venus 224,701 Tage\footnote{Williams 2005.}. 
Da die Erde sich ja auch 
um die Sonne bewegt (mit einer siderischen Umlaufzeit von 365,256 Tagen), folgen zwei 
gleiche Konstellationen --– wie etwa die untere Konjunktion --– mit einer Periode von 
583,924 Tagen, der sogenannten synodischen Periode. Wie man sofort erkennt, findet 
das wiederholte Ereignis zu einer anderen Jahreszeit statt, auch wird die Venus 
sich an einer anderen Stelle des Himmels vom Betrachter aus befinden. Um aus der 
Sicht eines Beobachters, welcher immer am selben Ort Positionen der Gestirne notiert, 
die Venus wieder etwa am gleichen Ort und zur gleichen Jahreszeit am Himmel zu finden, 
muss man also das kleinste gemeinsame Vielfache der synodischen Periode der Venus und 
des siderischen Jahres der Erde suchen. Dabei findet man, dass 8 Jahre (2922 Tage) 
fast genau 5 synodische Perioden der Venus entsprechen, der Unterschied beträgt 
2 Tage und 9,6 Stunden. Ein Beobachter findet die Venus also nach 8 Jahren 
näherungsweise am selben Tag im Kalender an derselben Stelle am Himmel wieder. 
Dieser 8-Jahreszyklus  war schon den Astronomen der Babylonier bekannt und ist 
in den Venustafeln des Ammi-saduqa aufgezeichnet worden\footnote{Bosanquet/Sayce 1879; 
Weir 1982. Eine allgemeine Übersicht bietet Ossendrijver 2008 mit weiterer Literatur, 
eine umfangreiche Bibliographie ist bei Walker/Galter/Scholz 1993 zu finden.}. 
Es ist allerdings kein 
exakter 8-Jahres-Zyklus, sondern es gibt von Periode zu Periode eine Verschiebung 
um den schon erwähnten Unterschied von etwa 2,4 Tagen bzw. fortlaufende Unterschiede 
bei markanten Stellungen wie zum Beispiel den nördlichsten und südlichsten 
Stellungen der Venus.

Da die Venus als Morgen- oder Abendstern nach Sonne und Mond zu den hellsten Gestirnen 
an unserem Himmel gehört, ist es naheliegend, nach beobachtbaren Extremen der Venusbahn, 
den Venuswenden im Aufgang und Untergang zu suchen. Die Venusbahn verläuft wie die 
aller Planeten in der Nähe der Ekliptik, d.h. wir erwarten die Venuswenden in der 
Nähe der Sonnenwenden. Eine Eigenschaft der Venusbahn ist noch erwähnenswert: die 
Bahn hat eine (heliozentrische) Neigung von 3,39$^\circ$ gegenüber der Ekliptik. Steht die 
Venus in ihrer größten Erdnähe, dann könnte dies theoretisch zu einer geozentrischen 
Breite der Venusbahn von 9,67$^\circ$ führen. Da aber die größte Erdnähe meistens nicht mit 
der größten (heliozentrischen) Breite der Venusbahn zusammenfällt, tritt dieser Wert 
nur sehr selten auf. Wir suchen also Venuswenden, die um einige Grad nördlicher als 
die nördliche Sonnenwende bzw. einige Grad südlicher als die südliche Sonnenwende 
liegen könnten.

\vspace{1em}
\noindent
Die Berechnungen der Venusephemeriden werden mit dem On-line Solar System Data 
Service des Jet Propulsion Laboratory durchgeführt\footnote{Giorgini, J.D., Yeomans, 
D.K., Chamberlin, A.B., Chodas, P.W., Jacobson, R.A., Keesey, M.S., Lieske, J.H., 
Ostro, S.J., Standish, E.M., Wimberly, R.N., ,,JPL's On-Line Solar System Data Service'', 
Bulletin of the American Astronomical Society, 28, 1158, 1996, siehe 
Website http://ssd.jpl.nasa.gov/?horizons.}.  Dieses Programm gestattet 
die Verwendung hochpräziser Bahndaten der Planeten, die Berücksichtigung der genauen 
Form des Erdglobus und der Lichtbrechung in der Atmosphäre, die vor allem bei 
horizontnahen Ephemeriden zu deutlichen Verschiebungen führt. Das verwendete 
Horizontkoordinatensystem hat im Norden den Azimutwinkel 0$^\circ$, im Osten 
90$^\circ$, im Süden 180$^\circ$ 
und im Westen 270$^\circ$.

Die Berechnungen der Venusephemeriden wurden für die Koordinaten des Tempels auf der 
Nordterrasse durchgeführt: geographische Breite 39,3100174$^\circ$ N, geographische 
Länge 36,9112600$^\circ$ O und Höhe über N.N.\ 1620~m. Das Programm verwendet für die 
Zeitangaben das julianische Datum, welches sich als Kalenderdatum im Julianischen 
Kalender für die Zeit der Hethiter ausgeben lässt, wohl wissend, dass dies nur 
eine Übertragung aus unserem Kulturkreis und nicht der damals gültige Kalender 
bei den Hethitern ist. Bei den im Folgenden enthaltenen Kalenderdaten gilt es 
auch noch zu berücksichtigen, dass aufgrund der Präzession der Erde sich die 
Jahreszeiten im Kalender verschieben. 

So fällt zum Beispiel die Sommersonnenwende im Jahre 1530 v.\ Chr.\ auf den 6.\ Juli. 
Ein Beobachter auf dem Dach des Tempels der Nordterrasse konnte diese am Horizont 
unter einem Azimut-Winkel von 57,67$^\circ$ beobachten. Allerdings ist zu beachten, dass 
sich im Osten und Süden der Stadt ein Bergrücken befindet. Die Horizontlinie liegt 
somit nicht in der Ebene des Tempels sondern deutlich höher. Berücksichtigt man 
eine Horizonthöhe von etwa 2$^\circ$, dann konnte die Sonnenwende unter 
59,84$^\circ$ beobachtet werden.

\begin{figure*}[t]
\begin{figs}
\begin{center}
\includegraphics[angle=0,width=1.05\textwidth]{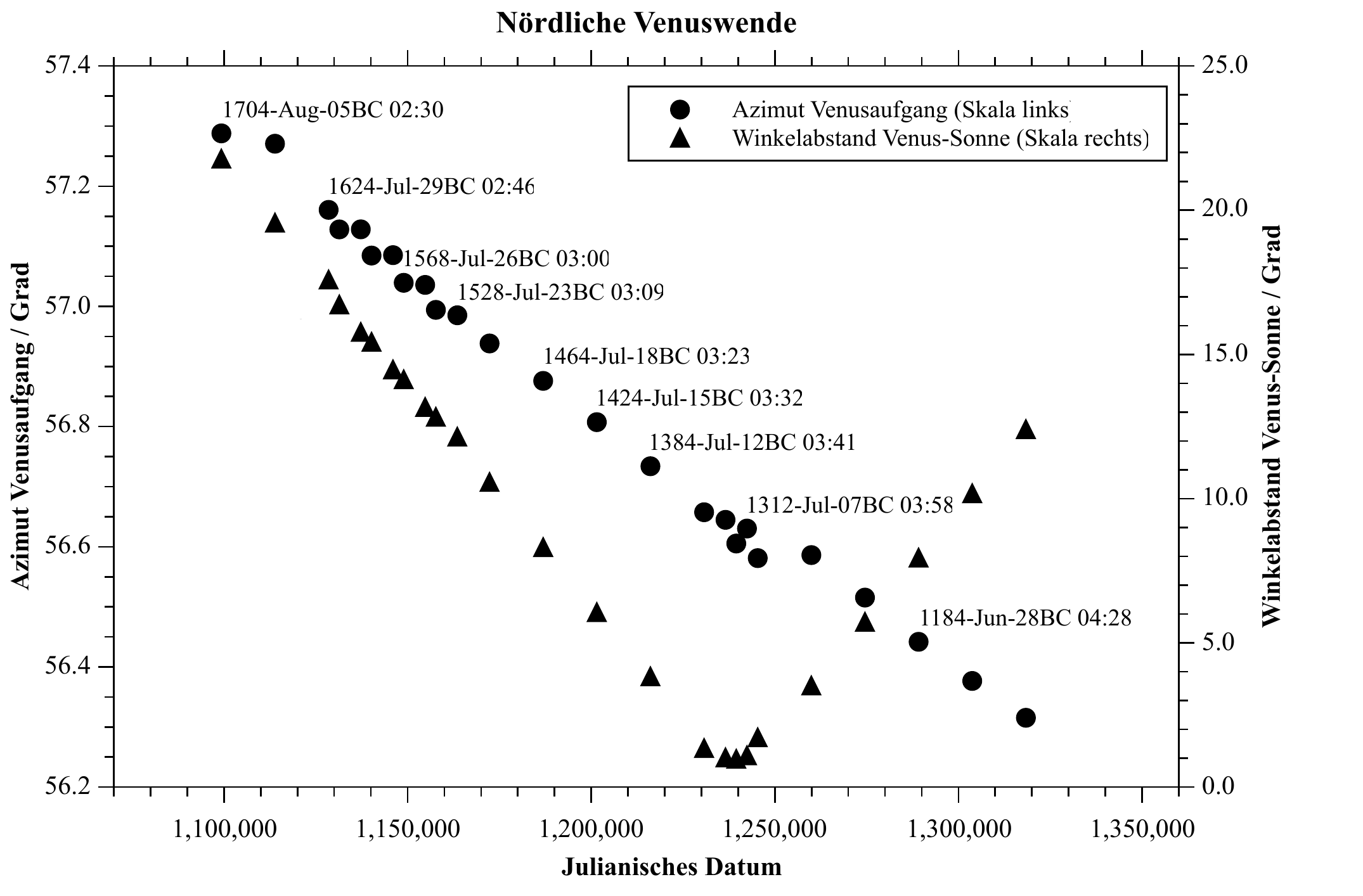}

\vspace*{0.5em}
Abb.\ 4: Nördliche Venuswende für einen ausgewählten Zyklus.\\
\parbox{\textwidth}{
Dargestellt ist in einer Acht-Jahresfolge des jeweils nördlichsten Venusaufgangs --- 
die sogenannte nördliche Venuswende --- vom 17.\ Jh.\ v.\ Chr.\ bis ins 12.\ Jh.\ v.\ Chr.\ 
(schwarze Punkte, linke Skala) sowie der Winkelabstand der Venus zur Sonne (schwarze Dreiecke, 
rechte Skala). Dies ist der einzige gefundene Zyklus mit beobachtbaren Venuswenden, 
die nördlicher als die Sonnenwenden liegen. Die Sichtbarkeit der Venus ist gegeben, 
wenn ihr Abstand zur Sonne größer als 5,4$^\circ$ ist. Vor dem Jahre 1312 v.\ Chr.\ 
steht die Venus bei ihrer Wende in diesem Zyklus vor der Sonne, sie zieht dann an der 
Sonne vorbei und geht erst nach der Sonne auf, d.h., sie ist dann nicht mehr als 
Morgenstern sondern als Abendstern sichtbar.
}
\end{center}
\end{figs}
\end{figure*}

Bei der Suche nach Venuswenden, die nördlicher als die Sonnenwenden liegen und 
beobachtbar sind, ist uns nur ein Zyklus aufgefallen (Abb.~4). Man sieht, 
dass in der Folge dieses Zyklusses die Venuswenden immer weiter nach Norden rutschen. 
Zwei Einschränkungen sind zu beachten: zum einen sind Venusauf- und Untergänge nur 
beobachtbar, wenn sie nicht zu dicht an der Sonne liegen. Als minimalen Abstand 
gibt C.\ Schoch 5,4$^\circ$ an\footnote{Schoch 1924  (online verfügbar: 
http://adsabs.harvard.edu/full/1924AN...222...275).}. 
Des weiteren liegt im dargestellten Zyklus die Venus 
bis zum Jahre 1312 v.\ Chr.\ vor der Sonne, ist also Morgenstern, für spätere Zeiten 
jedoch hinter der Sonne und wird zum Abendstern. Eine visuelle, nicht durch 
Hilfsmittel verstärkte Beobachtung des Venusaufganges am Taghimmel ist dann unmöglich. 
In diesem Zyklus finden wir also als nördlichste sichtbare Venuswende den Wert aus 
dem Jahre 1424 v.\ Chr.\ von 56,81$^\circ$, der allerdings nach Erbauung des Tempels liegt. 
Vorher käme dann nur der Wert von 1544 v.\ Chr.\ in Frage\footnote{Nach dem derzeitigen 
Stand der Untersuchungen wäre auch ein Errichtungsdatum 1544 v.\ Chr.\ für den Tempel 1 
durchaus möglich. Zwar spricht Vieles dafür, dass die übrige Stadt im wesentlichen 
in den 30er und 20er Jahren des 16.\ Jh.\ v.\ Chr.\  gebaut wurde (Fälldaten von 
Bauhölzern im Tempel 2/Gebäude C: 1529 +4/-7 v.\ Chr.; Bauhölzer aus zwei verschiedenen 
Stadttoren 1534 bzw. 1508 +4/-7 v.\ Chr.\ als jeweils letzter erhaltener Jahresring, 
d.h.\ Fälldatum einige Jahre später: Kuniholm/Newton 2002), es könnte mit dem Bau des 
Tempels 1 aber bereits vor der Anlage des Stadtmauerrings und des Tempels 2 begonnen 
worden sein. Kiefernbalken von Bauhölzern des Tempels 1 zeigen für den jüngsten 
erhaltenen Jahresring das Datum 1582 +4/-7 v.\ Chr. Die Rinde mit den letzten Jahresringen 
war abgebeilt worden. Es ist denkbar, dass nur ca.\ 38 Jahresringe bis zu dem potentiellen 
,,astronomischen'' Errichtungsdatum 1544 v.\ Chr.\ fehlen, es können aber auch rund 50 
Jahresringe fehlen, dann wäre der Tempel gemeinsam mit den übrigen Großbauten der 
Stadt entstanden.}, 
nämlich 57,00$^\circ$. Beide Ergebnisse liegen innerhalb eines Grades Abstand zur Sonnenwende. 
Die wirklich beobachteten Werte hängen dann noch von der Horizonthöhe ab und 
verschieben sich wie die Sonnenwende um etwas mehr als 2$^\circ$ Richtung Süden. Man 
erhält also 59$^\circ$ für die letzt mögliche Beobachtung der nordöstlichsten sichtbaren
Venuswende vor Erbauung des Tempels. Dies 
ist eine erstaunlich gute Übereinstimmung mit dem Wert von 59,1$^\circ$, dem Winkel der 
Längsachse des Tempels. Allerdings trifft dies auch auf die nördliche Sonnenwende zu.

\begin{figure*}[t]
\begin{figs}
\begin{center}
\includegraphics[angle=0,width=1.00\textwidth]{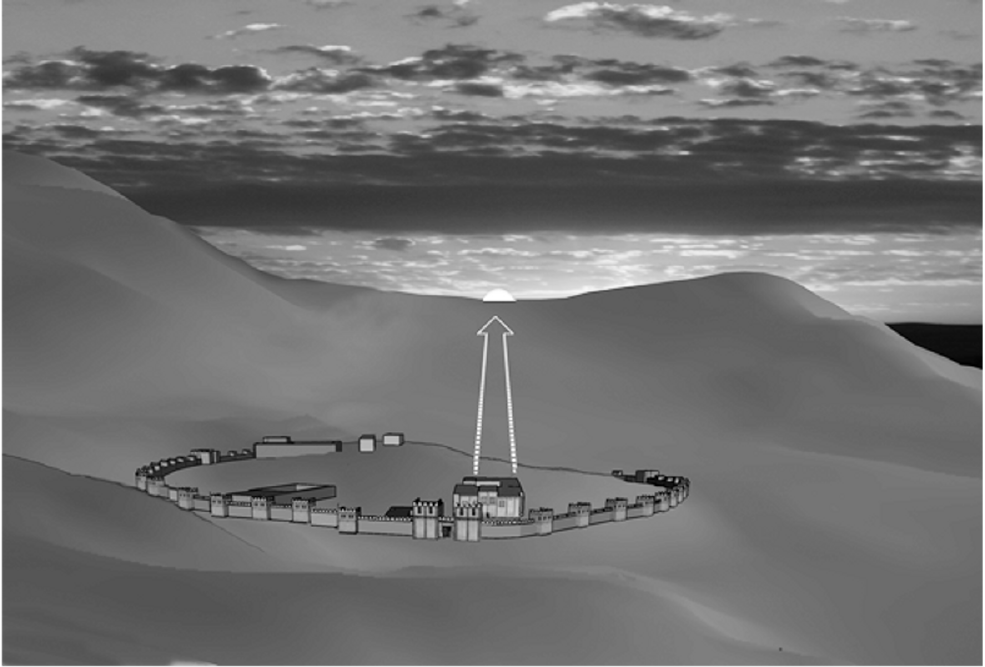}

\vspace*{1.5em}
\parbox{\textwidth}{
Abb.\ 5: Virtuelle Rekonstruktion der hethitischen Stadt Sarissa zum Zeitpunkt 
des Sonnenuntergangs am Tag der Wintersonnenwende. An diesem Datum verschwindet 
die Sonne genau in der Verlängerung der Längsachse des Tempels 1 hinter dem Horizont. 
Dieser Punkt liegt zudem am Boden eines kleinen Taleinschnittes.
}
\end{center}
\end{figs}
\vspace*{1em}
\end{figure*}

\vspace{1.5em}
\noindent
Blickt man in Verlängerung der Tempellängsachse nach Südwesten, so beträgt hier die 
Horizonthöhe (von der Geländeoberfläche aus betrachtet) lediglich ca.\ 1$^\circ$. Vom 
Tempeldach aus dürfte man ursprünglich sogar einen weitgehend freien Blick zu dem 
sich in mehreren Kilometern entfernten, auf etwa gleicher Höhe befindlichen Horizont 
gehabt haben. Allerdings war dort die Horizontlinie nicht gerade, denn exakt in 
Verlängerung der Tempellängsachse befindet sich ein beidseitig von Höhenzügen 
eingerahmtes kleines Tal (Abb.~5).

\begin{figure*}[t]
\begin{figs}
\begin{center}
\includegraphics[angle=0,width=1.14\textwidth]{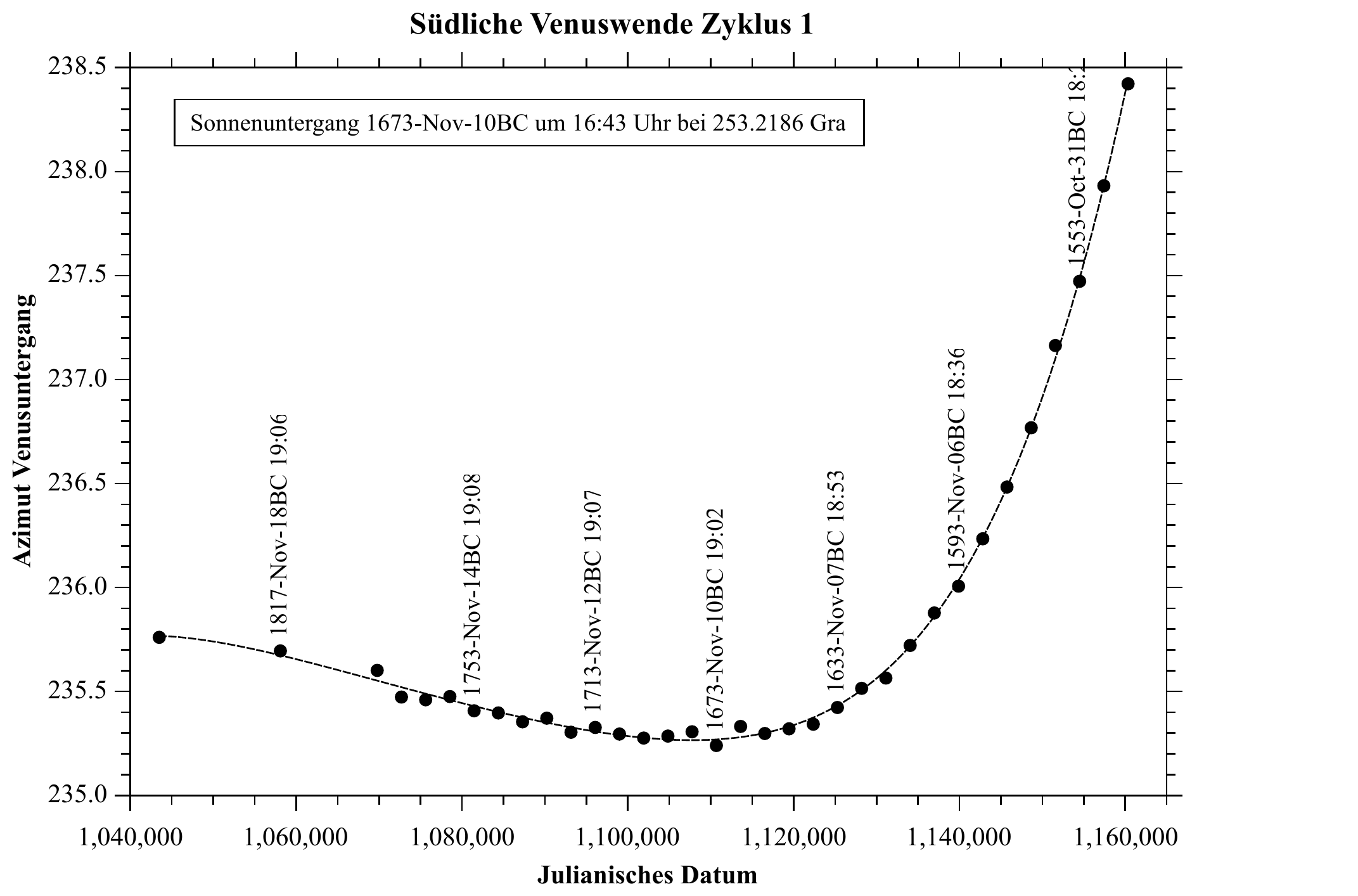}

\vspace*{0.5em}
\parbox{\textwidth}{
Abb.\ 6: Südliche Venuswende für einen Zyklus vor der Besiedlungszeit in Sarissa.
Dargestellt ist in einer Acht-Jahresfolge der jeweils südlichste Venusuntergang, 
die sogenannte südliche Venuswende vom 19.\ Jh.\ v.\ Chr.\ bis ins 16.\ Jh.\ v.\ Chr. 
Die Venuswende rutscht vom 18.\ November bis Ende Oktober vor, der südlichste Punkt 
dieser Folge liegt im Jahre 1673 v.\ Chr.\ bei 235,24$^\circ$.
}
\end{center}
\end{figs}
\vspace*{-1.5em}
\end{figure*}

Entsprechend wäre daran zu denken, dass nicht der Venusaufgang, sondern der Punkt 
des Untergangs des Planeten am Horizont den Orientierungspunkt für die Ausrichtung 
des Tempels abgab. Möglicherweise war für die Wahl des Bauplatzes sogar mit 
ausschlaggebend, dass von dort aus betrachtet die Venus nicht irgendwo am Horizont 
verschwand, sondern am Grund des eindrucksvoll von Berghängen flankierten Tälchens.

In diesem Fall kommt als Marke der südlichste Wendepunkt der Venus in Frage. 
Abb.~6 zeigt die südlichen Venuswenden für einen Zyklus vor der Besiedlungszeit, 
Abb.~7 einen weiteren während und nach der Besiedlungszeit von Sarissa. Diesmal 
sieht man sehr schön, wie beide Zyklen zu einer immer weiter im Süden liegenden 
Venuswende führen, einen Extremwert erreichen und dann wieder Wenden in weiter 
nördlichen Lagen auftauchen. In beiden Zyklen gibt es keine Sichtbarkeitseinschränkungen, 
in den Minima geht die Venus etwa 3 Stunden nach der Sonne unter, ist also als 
brillanter Abendstern sichtbar. Das Minimum vor der Erbauung des Tempels 
(Zyklus 1, Abb.~6) liegt bei 235,24$^\circ$, nach seiner Erbauung liegt es im 
Jahre 1430 v.\ Chr.\ bei 235,10$^\circ$ (Zyklus 2, Abb.~7). 
Beide Werte weichen um etwa 4$^\circ$ 
von der Tempelausrichtung ab (239,1$^\circ$ bei Blick nach Südwesten). 
Eine Horizonthöhe von 1$^\circ$ würde die Abweichung um 1,1$^\circ$ 
erhöhen, da die Sonne dann ja etwas später aufgeht und somit noch weiter südlich 
steht. Dies ist eine merkliche Abweichung und lässt doch Zweifel an einem 
astronomischen Bezug zu den Venuswenden aufkommen. 

\begin{figure*}[t]
\begin{figs}
\begin{center}
\includegraphics[angle=0,width=1.14\textwidth]{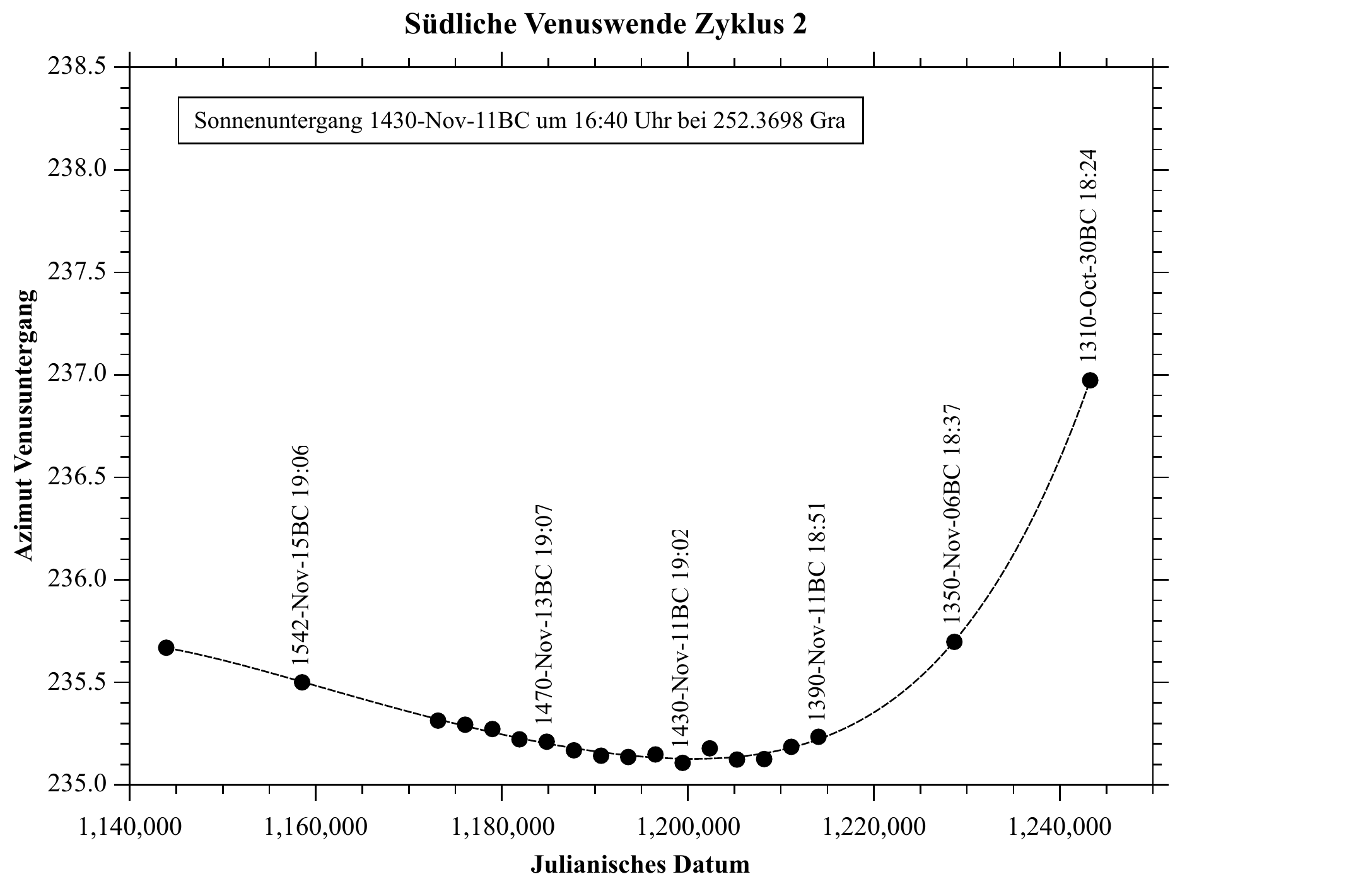}

\parbox{\textwidth}{
Abb.\ 7: Südliche Venuswende für einen Zyklus für die Zeit nach Erbauung der Stadt 
Sarissa. Dargestellt ist in einer Acht-Jahresfolge der jeweils südlichste Venusuntergang, 
die sogenannte südliche Venuswende vom 16.\ Jh.\ v.\ Chr.\ bis ins 14.\ Jh.\ v.\ Chr. 
Auch für diesen Zyklus rutscht die Venuswende von Mitte November bis Ende Oktober vor, der 
südlichste Punkt dieser Folge liegt im Jahre 1430 v.\ Chr.\ bei 235,10$^\circ$.
}
\end{center}
\end{figs}
\vspace*{-1.5em}
\end{figure*}

Es bleibt anzumerken, dass die Beobachtung solcher Planetenwenden nur möglich ist, 
wenn man mehrere Venuszyklen systematisch beobachtet und dokumentiert hat. Bei der 
über Jahrhunderte kontinuierlich betriebenen ,,Sternwarte'' von Babylon konnten solche 
Phänomene durchaus wahrgenommen, registriert und entsprechend für Vorhersagen 
genutzt werden. Derartiges ist aber für Sarissa kaum anzunehmen, zumal vor 
Errichtung des Tempels hier keine Stadt existierte, somit die notwendigerweise 
vor Ort durchzuführenden Langzeitbeobachtungen auch nicht erfolgen konnten. 
Ganz außergewöhnlich wäre solch ein astronomischer Bezug jedoch nicht. 
Neben den Astronomen in Babylon haben ja auch die mesoamerikanischen Kulturen 
detaillierte Beobachtungen der Planeten festgehalten. Bekannt ist ein solcher 
astronomischer Bezug einer ,,Sternwarte'' zu Venuswenden aus Mesoamerika. 
Vom Palast des Gouverneurs der Maya-Stadt Uxmal blickt man über die Pyramide 
Nohpat zum Aufgangspunkt der südlichen Venuswende\footnote{Aveni/ Hartung 1986.}.

Nicht von der Hand zu weisen ist jedoch auch ein Bezug zum Lauf der Sonne. 
Für das Jahr 1529 v.\ Chr.\ lässt sich der Untergangspunkt der Sonne am kürzesten 
Tag des Jahres, der Wintersonnenwende am 1.\ Januar 1529, auf 239,29$^\circ$ berechnen, 
eine sehr gute Übereinstimmung mit der Tempelachse (Abb.~5). Gleichwohl kommt 
natürlich auch der Aufgangspunkt der Sonne am Tag der Sommersonnenwende in Frage, 
der bei einer Horizonthöhe von 2$^\circ$ für die Blickrichtung nach Nordosten mit 59,84$^\circ$ 
gleichfalls eine erstaunlich gute Übereinstimmung zur Tempelausrichtung (59,1$^\circ$) 
zeigt, die kaum mehr als Zufall betrachtet werden kann.

\section*{Die Planungsachsen der Stadt}

Wenn wir nun nochmal einen Blick auf die Ausrichtung der Planungsachsen der Stadt werfen, 
kann sich der Eindruck einer sonnenorientieren Stadt vielleicht noch etwas erhärten. 
Warum haben die Erbauer eine Ausrichtung der Stadt um rund 45$^\circ$ verdreht zum Kreuz der 
Himmelsrichtungen gewählt? Die Himmelsrichtungen konnten mithilfe des Indischen Kreises 
bestimmt werden, die Ausrichtung des Tempels konnte nach Sonnenauf- und Untergangsmarken, 
vielleicht auch nach den Venuswenden, am Horizont geplant werden, was aber steckt hinter
der Idee, die Planungsachsen der Stadt an einer zu den vier Himmelsrichtungen
verlaufenden Diagonale auszurichten?

Man stelle sich einen Beobachter auf dem leeren zu planenden Gelände vor, der die 
Himmelsrichtung festgelegt hat und dann nach weiteren Hinweisen zur Orientierung 
sucht. Er findet, dass die Sonne jeden Tag unter einem schrägen Winkel aufgeht und 
auch unter dem gleichen Winkel wieder untergeht. Zur Tag- und Nachtgleiche ist der 
Auf- und Untergangswinkel der Sonnenbahn rechnerisch durch (90$^\circ$-geographische Breite), 
also 50,69$^\circ$ gegeben --- wenn man einfach nur 2 Ebenen (Horizont und Ekliptik) in 
sphärischer Geometrie betrachtet. Berücksichtigt man noch die Brechung des Lichtes 
in der Atmosphäre, dann werden vor allem die Beobachtungen in Horizontnähe sich 
anders darstellen. Es ergibt sich dann ein etwas kleinerer Winkel von 46,7$^\circ$, der 
nach einem Sonnenstand von einigen Grad über dem Horizont sich zunächst auf fast 50$^\circ$ 
vergrößert. Verfolgt man die Sonnenbahn beim Aufgang an anderen Tagen, so stellt man 
fest, dass der Winkel immer kleiner wird, je dichter man an die Tage der Sonnenwenden 
herankommt. Abb.~8 und 9 zeigen den Sonnenaufgang zur Sommersonnenwende und den 
Sonnenuntergang zur Wintersonnenwende, in beiden Fällen finden wir etwa 42$^\circ$ als 
Winkel zwischen Sonnenbahn und Horizont. In beiden Abbildungen eingezeichnet sind 
auch Linien, die einem Winkel von 45$^\circ$ der Sonnenbahn gegenüber dem Horizont 
entsprechen. Man erkennt leicht, dass in beiden Fällen einem Beobachter ohne 
optische Hilfsmittel die 45$^\circ$ Linien als Tangenten an die gekrümmten Kurven 
passend erscheinen, die Präzision einer Winkelbestimmung dieser Tangenten ist 
sicher nicht besser als 5$^\circ$ (Abb.~10). 

Das bedeutet, dass wir zu den Zeiten der Sonnenwenden, die vermutlich schon für 
die Orientierung des Tempels eine Bedeutung gehabt haben, Schräglagen der 
Sonnenbahn für den Aufgang und Untergang finden, die sehr gut zur verdrehten 
Orientierung der Planungsachsen der Stadt passen.

\begin{figure*}[t]
\begin{figs}
\begin{center}
\includegraphics[angle=0,width=1.10\textwidth]{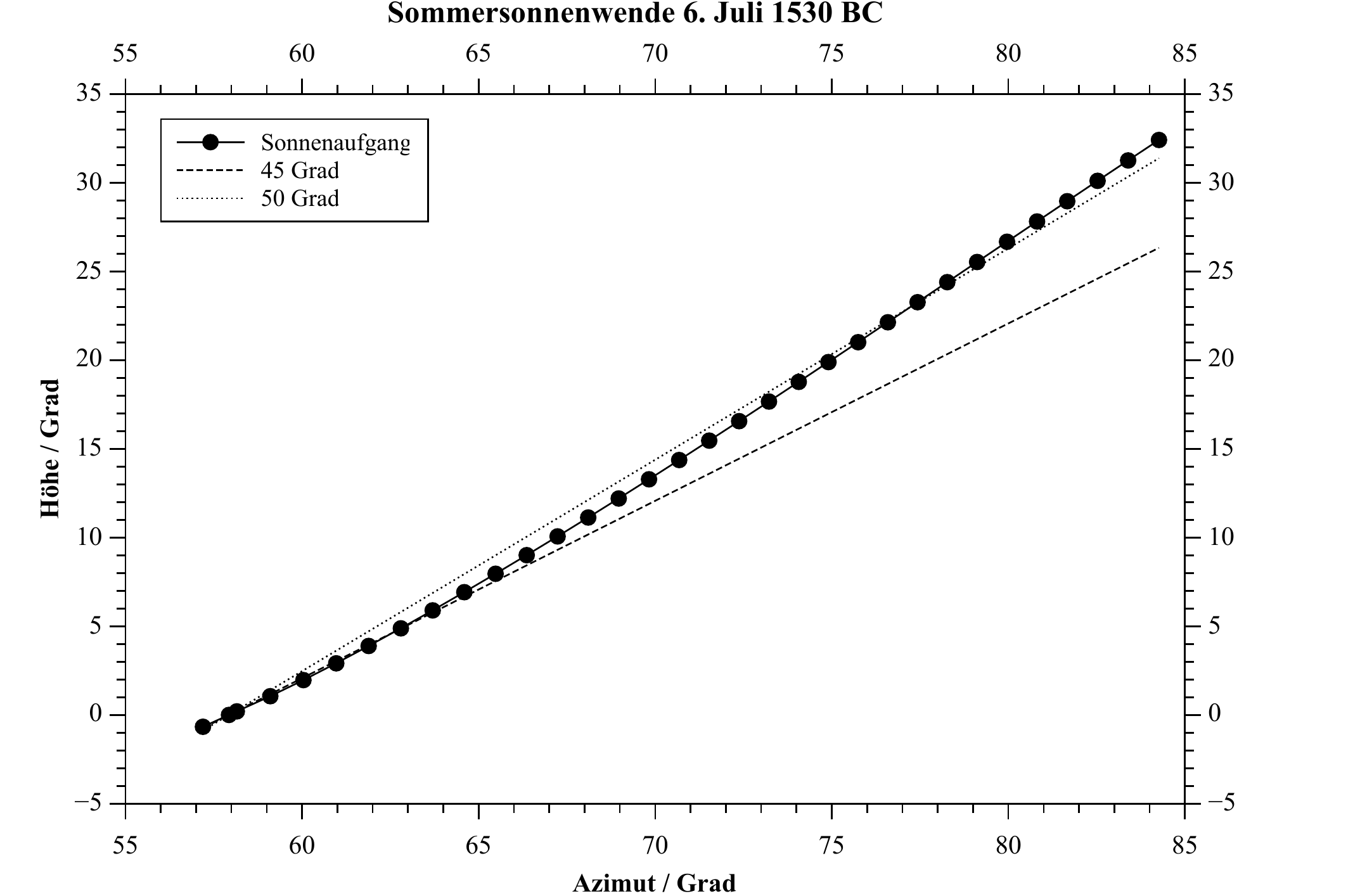}

\vspace*{1em}
\parbox{\textwidth}{
Abb.\ 8: Sonnenaufgang zur Sommersonnenwende am 6. Juli 1530 v.\ Chr.
Der Sonnenaufgang erfolgte für die geografische Breite von Sarissa bei einem 
Azimut von 57,67$^\circ$ (Norden ist 0$^\circ$) bei flachen Horizont. Der Tagbogen der 
Sonne ist eine gekrümmte Linie, die unter einem Winkel von 42,1$^\circ$ aus dem Horizont 
aufsteigt und sich dann stärker nach oben krümmt. Bei einer Höhe von 5$^\circ$ zeigt die 
Verbindungslinie zum Aufgangspunkt einen Winkel von 45$^\circ$ (die gestrichelte Linie 
in der Abbildung). Zur Abschätzung der Genauigkeit einer solchen Winkelbestimmung 
ist zusätzlich die punktierte Linie bei 50$^\circ$ eingezeichnet.
}
\end{center}
\end{figs}
\end{figure*}

\begin{figure*}[t]
\begin{figs}
\begin{center}
\includegraphics[angle=0,width=1.10\textwidth]{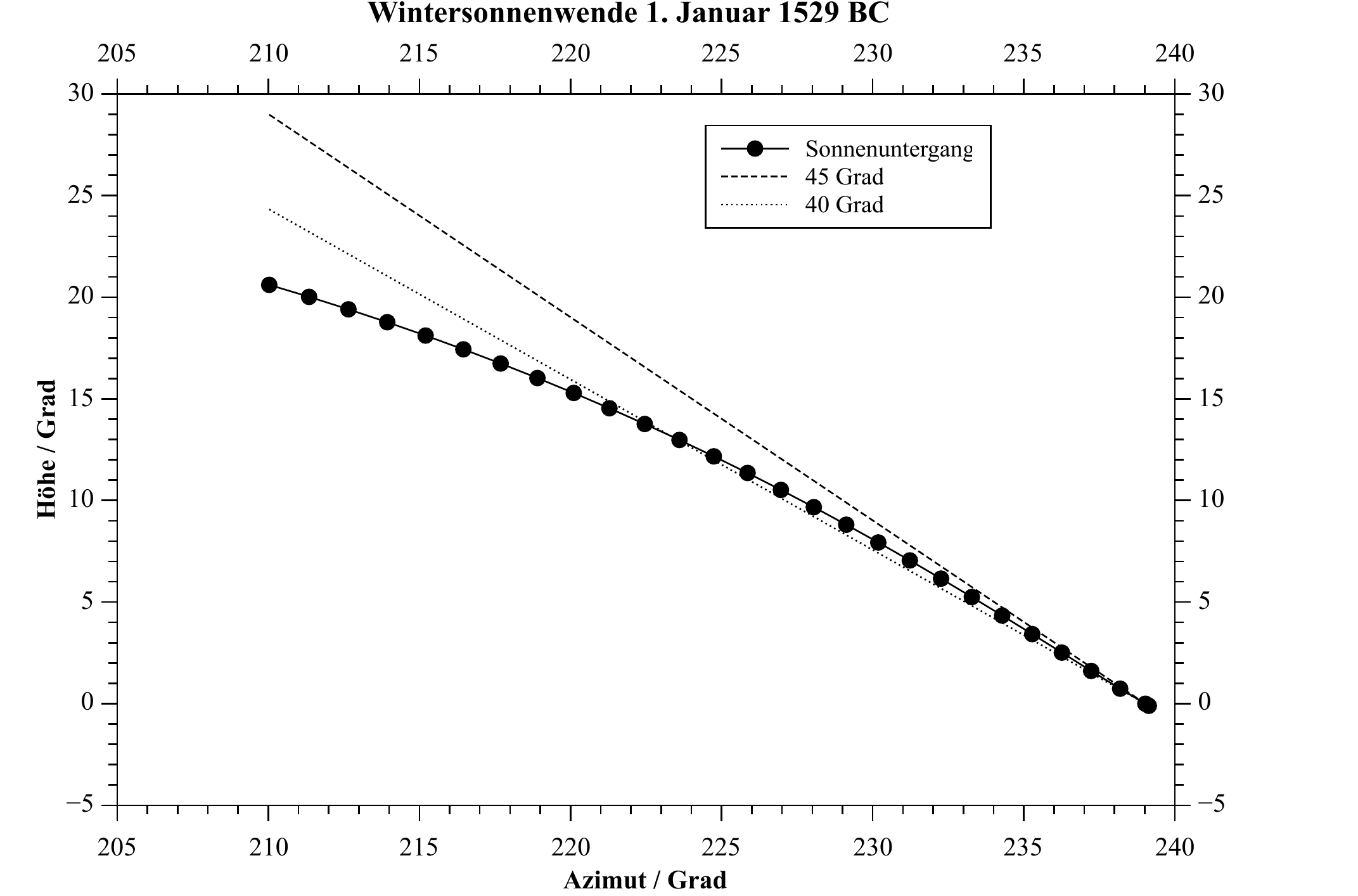}

\vspace*{1em}
\parbox{\textwidth}{
Abb.\ 9: Sonnenuntergang zur Wintersonnenwende am 1.\ Januar 1529 v.\ Chr. Der Sonnenuntergang 
erfolgt für die geografische Breite von Sarissa bei einem Azimut von 239,29$^\circ$ 
(Norden ist 0$^\circ$). 
Der Tagbogen der Sonne ist eine gekrümmte Linie, die unter einem Winkel von 41,5$^\circ$ aus dem 
Horizont aufsteigt, dann zunächst leicht auf 42,5$^\circ$ bei 5$^\circ$ Höhe ansteigt und sich 
dann stärker nach unten krümmt. Die gestrichelte Linie in der Abbildung hat eine 
Neigung von 45$^\circ$, die punktierte eine von 40$^\circ$. Die anfänglich steigende Neigung 
ist ein Effekt der Brechung der Atmosphäre.
}
\end{center}
\end{figs}
\end{figure*}

\section*{Sarissa --- eine Stadt mit astronomischem Bezug?}

Während der rund 300-jährigen Nutzungszeit des Tempels war jeweils nur an einem 
Tag des Jahres (den wir heute 21.\ Juni nennen) eine exakte Parallelität der ersten 
Sonnenstrahlen am Morgen mit den Außenwänden des Tempels zu beobachten. Die Nordwand 
warf nur an diesem Tag keinen Schatten. Analog zeigte sich das gleiche Phänomen genau 
ein halbes Jahr später zur Zeit der Wintersonnenwende: Nur an diesem Tag war die 
Südfassade des Tempels bis zum Sonnenuntergang vom Sonnenlicht beschienen und lag 
nicht wie sonst das Jahr über ab dem Nachmittag im Schatten. Die Bestimmung zweier 
wesentlicher Fixpunkte im Jahreslauf konnte somit an den Längsseiten des Tempels 
leicht vorgenommen werden. Weitere Kalenderdaten waren dann durch Zählung der 
Tage ab diesen Fixpunkten unschwer genau zu bestimmen. Die Beobachtungen können 
ebenso von dem zweifellos begehbaren Dach aus erfolgt sein, das als Flachdach mit 
Erdabdeckung zu rekonstruieren ist. Auch eignete sich der Innenhof oder der 
Sonneneinfall durch Fenster für derartige Beobachtungen. Die orthogonale 
Grundrissstruktur bedingt, dass 
sämtliche Mauerzüge der Anlage auch die 
sonnenwendpunktebezogene Orientierung aufwiesen bzw.\ im rechten Winkel zu dieser 
Achse laufen.

Man würde den Erbauern aber sicher nicht gerecht, klassifizierte man die Anlage 
nun als ,,Kalenderbau'', dessen Funktion die Bestimmung der Sonnenwenden gewesen sei. 
An welchem Punkt des Horizontes die Sonne an den Extrempunkten ihrer scheinbaren 
Bahn im Verlauf des Kalenderjahres auf- bzw. unterging dürfte den erfahreneren 
Bewohnern Sarissas auch so bekannt gewesen sein. Mehr noch wird hier die 
Ausrichtung der Längsachse des Tempels eine symbolische Bedeutung gehabt haben. 
Der Tempel spiegelt in seiner Orientierung etwas Überirdisches, er spiegelt den 
Sonnenlauf als Teil der kosmischen Ordnung.

Ist aber damit bereits auch die Frage entschieden, ob für den Tempel primär ein 
solarer Bezug oder ein Bezug zur Venus intendiert war? Auszuschließen ist 
Letzteres keinesfalls, ist doch – wie erwähnt – eine durchaus gute Übereinstimmung 
des Aufgangspunktes der Venus an ihrem nördlichsten Wendepunkt und eine grobe 
Übereinstimmung des südlichsten Untergangspunktes mit der Ausrichtung der 
Längsachse des Tempels gegeben. Von anderen Orten, an denen man Langzeitbeobachtungen 
der Gestirne vorgenommen hatte, dürfte man die Kenntnis erworben haben, 
dass der südliche Wendepunkt der Venus etwa mit dem Untergangspunkt der Sonne am 
Tag der Wintersonnenwende zusammen fällt\footnote{Diese Erkenntnis war der Astronomen 
in Babylon in dieser Epoche bereits seit mehreren Generationen geläufig (Pingree 1993). 
Nachweislich gelangte astronomisches Wissen aus Babylon zu den Hethitern. Ob dies 
bereits mit dem Zug Mursilis I nach Babylon erfolgte (diskutiert werden derzeit noch 
die Daten 1595, 1531 und 1523 v.\ Chr.) oder erst später, ist unklar. Sichere Belege 
liegen ab mittelhethitischer Zeit vor (Koch Westenholz 1993). Es ist aber durchaus 
davon auszugehen, dass die in Sarissa tätigen hethitische Stadtplaner und Baumeister 
Zugang zu diesem Wissen hatten.}. Wollte man also den alle acht Jahre 
etwa an derselben Stelle beobachtbaren Untergangspunkt der Venus bestimmen, 
so war nur der Tag der Wintersonnenwende abzuwarten und man wusste, wo in 
etwa auch die Venus an ihrem südlichen Wendepunkt hinter dem Horizont verschwinden 
würde. Für einen möglichen Ištar-Tempel kann dieser Punkt durchaus von Bedeutung gewesen sein.

\begin{figure*}[t]
\begin{figs}
\begin{center}
\includegraphics[angle=0,width=1.0\textwidth]{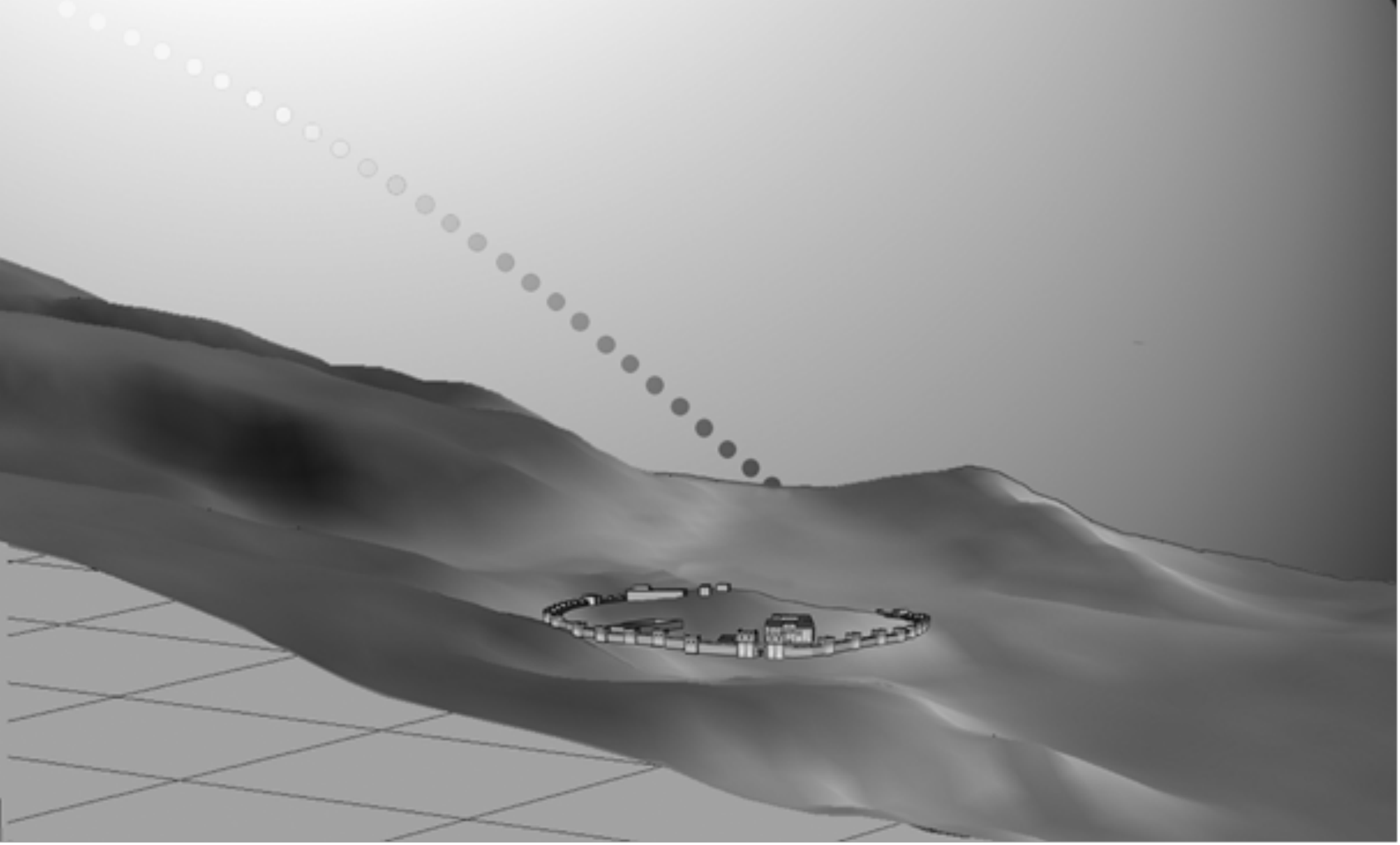}

\vspace*{1em}
\parbox{\textwidth}{
Abb.\ 10: Virtuelle Rekonstruktion der hethitischen Stadt Sarissa in ihrer Umgebung mit 
Darstellung der Sonnenbahn und ihres Untergangspunktes exakt in der Verlängerung der 
Längsachse des Tempels 1 am Tag der Wintersonnenwende. Der Neigungswinkel wird mit etwa 
45$^\circ$ wahrgenommen, dem Winkel, mit dem Stadt-Planungsachsen gegenüber den 
Haupt-Himmelsrichtungen verschoben sind.
}
\end{center}
\end{figs}
\vspace*{1em}
\end{figure*}

\vspace{1em}
\noindent
Falls tatsächlich ein Zusammenhang zwischen der Orientierung des Bauwerks und dem 
Lauf der Venus bestanden haben sollte, so wäre zu erwarten, dass sich dieser 
8-Jahres-Zyklus des Planeten im Kultkalender wiederspiegelt. Ein spezielles 
Ištar- bzw. Anzili-Ritual ist für Sarissa bislang nicht überliefert\footnote{Im Rahmen 
des wohl alljährlich stattfindenden Frühjahrsfestes, zu dem auch der Großkönig 
aus der Hauptstadt anzureisen hatte, wurde zwar sowohl der Wettergott von 
Sarissa wie auch Anzili und eine Schutzgottheit verehrt (Wilhelm 1997:10-18), 
es ist aber sicher damit zu rechnen, dass es darüber hinaus noch weitere, spezifische 
Rituale gab.}. Auch im 
Rahmen der sonst insgesamt sehr reichen schriftlichen Überlieferung zu 
hethitischen Kulthandlungen spielt die Zahl 8 keine besondere Rolle. Auffällig 
häufig sind hingegen die Belege für die Zahl 9 im hethitischen Schrifttum, 
insbesondere den Ritualtexten. Diesem Umstand widmete  erstmalig O.R.\ Gurney 1978 
eine Studie, die jüngst N.\ Oettinger aufgriff und erweiterte\footnote{Oettinger 2008.}. 
Beide arbeiteten 
überzeugend heraus, dass der Zahl 9 eine herausragende symbolische Bedeutung zukam. 
Vielfach wird in den Texten die Anzahl von Opfertieren mit 9 angegeben, es finden 
sich aber auch explizit Erwähnungen dieser Zahl im Zusammenhang mit einem Zyklus 
für Kulthandlungen:  Beispielsweise fanden Teile des bis in althethitische Zeit 
zurückgehenden Purulliya-Neujahrsfestrituals  nicht jährlich statt, sondern 
,,im neunten Jahr''\footnote{Haas 1994: 698. Der Zweck des Festes ist ,,primär eine 
Art der Welterneuerung – eine Verjüngung oder Regeneration der im Laufe der 
vergangenen Jahre verbrauchten Kräfte des Kosmos ...'' (Haas 1994: 679). Es ist 
zu erwarten, dass in einem solchen Kontext die drei wichtigsten kosmischen Zyklen, 
die der Sonne, des Mondes und der Venus, eine Rolle spielten.}. 
Diese Angabe ist entweder so zu verstehen, dass nach dem Ablauf 
von acht Jahren die Rituale durchzuführen waren oder man wollte das damit 
ausdrücken, was wir  heute als ,,alle acht Jahre'' bezeichnen würden: In einem 
Zahlensystem, dass die ,,Null'' nicht kannte (wie dem hethitischen) musste ein 
8-Jahreszyklus mit ,,neun'' ausgedrückt werden\footnote{Reste dieser Denkweise 
finden sich noch im heutigen Sprachgebrauch: Ein acht-jähriges Kind befindet sich 
im neunten Lebensjahr. Im Verständnis anderer Kulturen ist dieses Kind neun 
Jahre alt.}. Der Venus-Zyklus wäre somit 
von den Hethitern eher mit der Zahl ,,Neun'' als mit ,,Acht'' verbunden worden\footnote{Es 
ist auffällig, dass gerade auch in Ištar-Ritualen bei bestimmten Handlungen (Wasser-Schöpfen) 
die Zahl Neun genannt wird: Wegner 1981: 145 f oder neun Brote geopfert werden: 
Wegner 1981:139.}.  
Es ist daher durchaus denkbar, dass angesichts der zentralen Bedeutung der 
Ištar und ihrer verschiedenen Erscheinungsformen im hethitischen Pantheon die 
in religiösen Texten so häufig vorkommende Zahl 9 in Wirklichkeit von dem 
Venus-Zyklus abgeleitet ist\footnote{In diesem Zusammenhang könnte auch eine Passage 
in der Festbeschreibung CTH 591 Ia.A.(KBo 17.88 III, 10') sein. Hier heißt es 
im Kontext mit Wünschen für ein langes Leben des Königspaares ihnen mögen ,,9mal tausend, 
9mal doppelt, 9mal Jahre der Herrin'' gegeben sein. Der Ausdruck: 
9-\textsl{an} GAŠAN$^{\mbox{\tiny\textsl{TI}}}$   MU$^{\mbox{\tiny\U{H}I.A}}$-uš  ,,9mal Jahre der Herrin'' könnte auf den Ištar-/ Venus-Zyklus zu 
beziehen sein, da Ištar vielfach mit ,,GAŠAN/Herrin'' bezeichnet wurde (Wegner 1981: 33). 
Wie sich aus phonetischen Komplementen ergibt, wurde das Keilschriftzeichen GAŠAN in 
hethitischen Texten oftmals als Šaušga (der hurritischen Form für Ištar) gelesen 
(van Gessel 1998: 385). Die Verbindung zu dieser konkreten Göttin ist somit sehr eng. 
Zu der o.g. Textstelle siehe auch Klinger 1996: 318 f. 345 mit ausführlichem Kommentar 
sowie Haas 1994:194. Nicht unerwähnt sollte bleiben, dass in demselben Text nur 10 
Zeilen später zudem von dem Berg Sarissa die Rede ist (Klinger 1996: 321). Bemerkenswert 
ist ferner, dass in einem hethitischen Bauritual mit der Beschreibung der bei der Gründung 
eines Tempels auszuführenden Handlungen  explizit neun Opferdeponierungen vorzunehmen 
waren, bei denen u.a. neun Türminiaruren zu vergraben waren (Haas 1994;252-256).}. 

\newpage
Als ein weiteres Beispiel sei die späthethitische (luwische) Inschrift aus 
Hisarcık genannt, in der es  im Kontext eines Rituals für der Berg(gott) 
Harhara (Erciyes Dağ bei Kayseri) heißt: ,,... Und wenn das neunte Jahr eintritt, 
so [opfere] ich Dir neunmal (jeweils eine) \textsl{irwa}-Gazelle. ...''\footnote{E.Rieken (Marburg) 
danken wir für diese Übersetzung, die auf der Umschrift von Hawkins 2000:483 basiert: 
§ 3  9\textsl{-ti-sa-ha-wa/i-ti}[-\textsl{i}$^{\mbox{\tiny\textsl{?}}}$] 
ANNUS-\textsl{sa}$_4$-\textsl{si-sá-}' {\big \vert} REL-\textsl{ti} {\big \vert} \textsl{ta-i}
\ \ § 4 \ \ {\big \vert} \textsl{wa/i-tu-u}  \textsl{9-ta} {\big \vert} 
\textsl{i+ra/i-wa/i-ti} {\big \vert} {\big \vert} \ldots \textsl{wa/i?-ha}. 
Hawkins übersetzt diese Passage ,,And when(?) the ,,year's ninth” comes, . .”}. 
Diese Inschrift 
ist in diesem Zusammenhang auch besonders aufschlussreich, weil sie klar eine 
Verbindung zwischen der Zeitangabe ,,9'' und der Anzahl von 9 Opfertieren herstellt. 
Damit wird es noch wahrscheinlicher, dass sich auch sonst Nennungen von neun 
Opfergaben letztlich auf den kosmischen Rhythmus beziehen.

Wie dem auch sei --- ein Himmelsbezug dürfte bereits beim Auspflocken des 
Bauplatzes für den Tempel I in Sarissa eine Rolle gespielt haben. Nicht nur 
Baugrund, Hangneigung, Nachbarbebauung oder andere ,,terrestrische'' Kriterien 
waren für die Wahl der Längsachse ausschlaggebend, sondern wohl hauptsächlich 
der Gedanke, dass sich ein Aspekt der kosmischen Ordnung in dem Sakralbau 
wiederfinden sollte.

Besonders interessant ist der Gedanke, dass zum einen die Sonne, der hellste 
und durch sein Licht und seine Wärme auch dominierende Himmelskörper für die 
Bestimmung der Himmelsrichtungen mit dem Indischen Kreis und dann auch für 
den verdrehten Grundplan der Stadt Pate gestanden haben kann und zum anderen 
dann die Venus für die in diesem nach kosmischen Gesetzen geordneten Grund 
zur Hervorhebung des Ištar-Tempels beigetragen haben könnte.

\vspace{1cm}

\section*{\textup{\textsc{Bibliographie}}}
\begin{bib}
A.F.\ Aveni und H.\ Hartung 1986: Maya City Planning and the Calendar. Transactions\\ of
the American Philosophical Society 76, 1. 

\vspace{0.5em}
\noindent
R.H.M.\ Bosanquet and A.H.Sayce 1879: The Babylonian Astronomy.  Monthly Notes\\ of 
the Royal Astronomical Society 40, 565.

\vspace{0.5em}
\noindent
B.H.I.\ van Gessel 1998: Onomasticon of the Hittite Pantheon. Handbuch der Orientalistik I,33, Leiden/ New York/ Köln.

\vspace{0.5em}
\noindent
A.\ Goetze 1950: In: J.B.Pritchard (Hrsg.), Ancient Near Eastern Texts Relating to the Old
Testament, Princeton, 356 ff.

\vspace{0.5em}
\noindent
O.R.\ Gurney 1997: The Symbolism of 9 in the Babylonian and Hittite Literature. Journal of the Department of English. University of Calcutta 14, 27-31.

\newpage
\noindent
V.\ Haas 1994: Geschichte der hethitischen Religion. Handbuch der Orientalistik I, 15, Leiden, New York, Köln.

\vspace{0.3em}
\noindent
J.D.\ Hawkins 2000: Corpus of Hieroglyphic Luwian Inscriptions I. Inscriptions of The Iron Age 2, Berlin, New York.

\vspace{0.3em}
\noindent
J.\ Klinger 1996: Untersuchungen zur Rekonstruktion der hattischen Kultschicht. Studien zu den Boğazköy-Texten 37, Wiesbaden.

\vspace{0.3em}
\noindent
P.I.\ Kuniholm und M.\ Newton 2002: Dendrochronological Investigations at Kuşaklı-Sarissa. In: A.Müller-Karpe et al., Untersuchungen in Kuşaklı 2001. MDOG 134, 339-342.

\vspace{0.3em}
\noindent
A.\ Müller-Karpe 2000: Die Akropolis der hethitischen Stadt Kuşaklı-Sarissa. Nürnberger Blätter zur Archäologie 16, 1999/200, 91-110.

\vspace{0.3em}
\noindent
A.\ Müller-Karpe 2009a: Šarišša.\ B.\ Archäologisch. In: M.P.\ Streck et al. (Hrsg.) Reallexikon der Assyriologie und Vorderasiatischen Archäologie 12,1./.2. Šamu\U{h}a – Schild, 62-64, Berlin, New York.

\vspace{0.3em}
\noindent
A.\ Müller-Karpe 2009b: The Rise and Fall of the Hittite Empire in the Light of Dendroarchäological Research. In: St..W.Manning and M.J.Bruce (Hrsg.) Tree-rings, Kings and Old World Archaeology and Environment. Papers Presented in Honor of Peter Ian Kuniholm, Oxford, Oakville. 

\vspace{0.3em}
\noindent
P.\ Neve 1999: Die Oberstadt von \U{H}attuša. Die Bauwerke I. Die Bebauung im zentralen Tempelviertel. Boğazköy-\U{H}attuša XVI, Berlin.

\vspace{0.3em}
\noindent
M.\ Ossendrijver 2008: Astronomie und Astrologie in Babylonien. In: J.\ Marzahn, G.\ Schauerte (Hrsg.) Babylon – Wahrheit. Katalog zu Ausstellung: Babylon. Mythos und Wahrheit. Berlin 26.6. – 5.10.2008, 373-392, München.

\vspace{0.3em}
\noindent
N.\ Oettinger 2008: Zur Zahlensymbolik bei den Hethitern. In: A.\ Archi und R.\ Francia, VI.\ Congresso Internazionale di Ittitologia Roma, 5.-9.9.2005. Studi Micenei ed Egeo-Anatolici 50, 587-595.

\vspace{0.3em}
\noindent
D.\ Pingree 1993: Venus Phenomena In Enuma Anu Enlil. In: H.\ D.\ Galter (Hrsg.), Die Rolle 
der Astronomie in den Kulturen Mesopotamiens. Beiträge zum 3.\ Grazer Morgenländischen 
Symposion (23.-27.9.1991), 259–273, Graz.

\vspace{0.3em}
\noindent
W.\ Schlosser, G.\ Mildenberger, M.\ Reinhardt und J.\ Cierny 1997: Astronomische Ausrichtungen
im Neolithikum I. Ein Vergleich der böhmisch-mährischen Schnurkeramik und Glockenbecherkultur, Bochum.
Sterne und Steine. Eine praktische Astronomie der Vorzeit, Darmstadt.

\vspace{0.3em}
\noindent
C.\ Schoch 1924: Das Venus-Tablet Amizaduga $\left[\mbox{sic}\right]$. Astronomische Nachrichten 222, 27ff.

\vspace{0.3em}
\noindent
R.\ Stadelmann 1991: Die Ägyptischen Pyramiden. Vom Ziegelbau zum Weltwunder, 2.\ Aufl., Mainz.

\vspace{0.3em}
\noindent
C.\ B.\ F.\ Walker, H.\ D.\ Galter und B.\ Scholz 1993: Bibliography of Babylonian Astronomy 
and Astrology. In: H.\ D.\ Galter (Hrsg.), Die Rolle der Astronomie in den Kulturen 
Mesopotamiens. Beiträge zum 3.\ Grazer Morgenländischen Symposion (23.-27.9.1991), 407-449, Graz.

\vspace{0.3em}
\noindent
I.\ Wegner 1981: Gestalt und Kult der Ištar-Šawuška in Kleinasien. Alter Orient und Altes Testament 36. Hurritologische Studien 3, Neukirchen-Vluyn.

\newpage
\noindent
J.D.\ Weir 1982: The Venus Tablets: A Fresh Approarch. Journal for the History of Astronomy 13, 23-49.

\vspace{0.5em}
\noindent
U.\ Koch Westenholz 1993: Mesopotamian Astrology at Hattusas. In: H.\ D.\ Galter (Hrsg.), 
Die Rolle der Astronomie in den Kulturen Mesopotamiens. Beiträge zum 3.\ Grazer 
Morgenländischen Symposion (23.-27.9.1991), 231-246, Graz.

\vspace{0.5em}
\noindent
G.\ Wilhelm 1997: Keilschrifttexte aus Gebäude A. Kuşaklı-Sarissa 1,1, Rahden,Westf.

\vspace{0.5em}
\noindent
G.\ Wilhelm 2002: Die Keilschriftfunde der Kampagne 2002 in Kuşaklı. In: A.\ Müller-Karpe 
et al., Untersuchungen in Kuşaklı 2001, MDOG 134, 342-351.

\vspace{0.5em}
\noindent
G.\ Wilhelm im Druck: Die Lesung des Namens der Göttin \textsl{IŠTAR-li}. In: J.\ Klinger, 
E. Rieken (Hrsg.),  Gedenkschrift E.\ Neu.\ Studien zu den Boğazköy-Texten 52, Wiesbaden.

\vspace{0.5em}
\noindent
D.R.\ Williams 2005:  ,,Venus Fact Sheet'', NASA, http://nssdc.gsfc.nasa.gov/planetary/\\factsheet/venusfact.html. Retrieved 2007-10-12.
\end{bib}

\end{document}